# Controlling spin-orbit coupling to tailor type-II Dirac bands

Nguyen Huu Lam[†,#], Phuong Lien Nguyen[‡,#], Byoung Ki Choi[§,‖,#], Trinh Thi Ly[†], Ganbat Duvjir[†], Tae Gyu Rhee[‖], Yong Jin Jo[†], Tae Heon Kim[†], Chris Jozwiak[§], Aaron Bostwick[§], Eli Rotenberg[§], Younghun Hwang[$,*], Young Jun Chang[‖,*], Jaekwang Lee[‡,*], Jungdae Kim[†,*]

[†] Department of Physics, and EHSRC, University of Ulsan, Ulsan 44610, Republic of Korea

[‡] Department of Physics, Pusan National University, Busan 46241, Republic of Korea

[§] Advanced Light Source (ALS), E. O. Lawrence Berkeley National Laboratory, Berkeley, California 94720, United States

[‖] Department of Physics and Smart Cities, University of Seoul, Seoul 02504, Republic of Korea

[$] Electricity and Electronics and Semiconductor Applications, Ulsan College, Ulsan 44610, Republic of Korea

* Corresponding authors: younghh@uc.ac.kr; yjchang@uos.ac.kr; jaekwangl@pusan.ac.kr; kimjd@ulsan.ac.kr

**ABSTRACT**

$NiTe_2$, a type-II Dirac semimetal with strongly tilted Dirac band, has been explored extensively to understand its intriguing topological properties. Here, using density-functional theory (DFT) calculations, we report that the strength of spin-orbit coupling (SOC) in $NiTe_2$ can be tuned by Se substitution. This results in negative shifts of the bulk Dirac point (BDP) while preserving the type-II Dirac band. Indeed, combined studies using scanning tunneling spectroscopy (STS) and angle-resolved photoemission spectroscopy (ARPES) confirm that the BDP in the $NiTe_{2-x}Se_x$ alloy moves from +0.1 eV ($NiTe_2$) to –0.3 eV (NiTeSe) depending on the Se concentrations, indicating the effective tunability of type-II Dirac fermions. Our results demonstrate an approach to tailor the type-II Dirac band in $NiTe_2$ by controlling the SOC strength via chalcogen substitution. This approach can be applicable to different types of topological materials.

Quantum materials have been intensively investigated to understand their intriguing topological properties, which have the potential for utilization in future applications. A representative example of such quantum materials is a topological insulator. This material behaves as an insulator in bulk, but it has conducting surface states protected by non-trivial symmetry. Symmetry-protected Dirac fermions on the surface open an exciting opportunity to develop novel spintronic devices and transistors without dissipation.[1,2] Topological insulators are also suggested to host Majorana fermions when superconductivity is induced on the symmetry-protected surface states.[3] Topological materials can be classified into different types depending on the shape of the Dirac bands and symmetry configurations.[4] Type-I Dirac semimetals show isolated Dirac points with linear Dirac cones.[5–7] On the other hand, type-II Dirac semimetals with broken Lorentz invariance hold strongly tilted anisotropic Dirac cones near the Fermi level ($E_F$).[8,9]

Lately, type-II Dirac semimetals have attracted extensive interest due to their rich topological phenomena and potential applications. Indeed, strongly tilted Dirac cones give rise to distinct properties compared to those of type-I Dirac semimetals, such as chiral anomalies[10,11] and anisotropic magnetoresistance.[8] In addition, type-II Dirac semimetals with a finite density of states near Dirac points might provide a model system to realize topological superconductivity, which could host Majorana fermions applicable for topological quantum computing.[3,12–16]

Extensive theoretical and experimental investigations have been reported showing that group-10 transition-metal dichalcogenides (TMDs), such as $PtSe_2$, $PtTe_2$, and $PdTe_2$, hold type-II Dirac fermions.[17] Access to the Dirac fermions in these TMDs, however, is quite limited because the Dirac points are located far away from the $E_F$.[17–19] Therefore, the physical signature of relativistic particles is obscure because the electronic properties are mainly governed by the electrons at the $E_F$. Several efforts have been made to resolve this issue in type-II Dirac materials. Fei *et al.* showed that the $E_F$ can be tuned close to the bulk Dirac point (BDP) in $IrTe_2$ by doping with Pt.[20] On the other hand, it was reported that $NiTe_2$, another group-10 TMD, holds the BDP slightly above the $E_F$ by 70 – 100 meV, implying an ideal type-II Dirac system.[17,21,22] Then, the additional tunability of $E_F$ in $NiTe_2$ will greatly enhance the possibility of applications utilizing type-II Dirac fermions.

Here, we succeed in demonstrating systematic control of spin-orbit coupling (SOC) to tailor the type-II Dirac band in $NiTe_2$ via Se substitutional alloying. Both electronic and topological properties of $NiTe_2$ are closely associated with Te atoms because the Te $p$ states are located near the $E_F$ and the heavier Te atoms drive the type-II Dirac character. Using first-principle DFT calculations, we show that the Se substitution can effectively tune the SOC strength in $NiTe_2$ and form the stable $NiTe_{2-x}Se_x$ alloy even though the $1T$-phase $NiSe_2$ is unstable. Indeed, high-quality $NiTe_{2-x}Se_x$ alloys are synthesized. Combined studies of STS and ARPES confirm that the BDP in the $NiTe_{2-x}Se_x$ alloy moves from +0.1 eV ($NiTe_2$) to –0.3 eV (NiTeSe), depending on the amount of Se substitution, while preserving its type-II Dirac characteristics. The theoretical variations of BDP agree well with the STS and ARPES measurements.

**RESULTS AND DISCUSSION**

The octahedral prismatic $1T$-$NiTe_2$, schematically represented in Figure 1a, crystallizes in the $CdI_2$-type structure with the space group $D_{3d}^3$. Each slab of $1T$-$NiTe_2$ is composed of three layers, where the central Ni layer is sandwiched between two Te sublayers. $NiTe_2$ slabs are stacked together by weak van der Waals interactions, allowing easy mechanical exfoliation.

Band inversion driven by SOC plays an essential role in determining the topological characteristics. In this regard, it is very important to understand the role of heavy chalcogen elements when tailoring the topological properties of NiTe$_2$. Figure 1b shows high symmetry paths within the Brillouin zone of bulk NiTe$_2$. The Γ-A path (D-point) has the $C_{3v}$ spatial symmetry and the $C_3$ rotation, while the Γ-point and A-point have i$C_{3v}$ symmetry due to the inversion symmetry.

As shown in Figure 1c, without considering the SOC, the A$_1$-band and doubly degenerate E-band cross below the $E_F$ due to the different irreducible representations, resulting in a symmetry-protected BDP point (marked by the red circle in Figure 1c). The A$_1$-band is dominated by Te $p_z$, while the E-band is dominated by Te $p_x$, $p_y$ (see Figure S1, supporting information (SI)). Since the $p_z$-derived A$_1$-band is more dispersive than the $p_x$, $p_y$-derived E-band (shown in Fig.1c), the Dirac cone is strongly tilted along the $k_z$ direction, producing type-II Dirac fermions. When SOC is considered, the E-band splits into the $\Delta_4$-band and doubly degenerate $\Delta_{5,6}$-band (see Figure 1d). The $\Delta_{5,6}$-band moves above the $E_F$, while the $\Delta_4$-band moves below the $E_F$ near the Γ-point; these shifts are related to the strength of SOC ($\Delta E_{SOC}$). Accordingly, the symmetry-protected type-II BDP moves to about 0.1 eV above the $E_F$ and $k_z = \pm 0.35 c^* (c^* = 2\pi/c)$ (near the A-point), as shown in Figure 1d.[23–25]

In comparison with NiTe$_2$, the imaginary frequency modes in the phonon spectra (SI Figure S2) indicate that 1$T$-phase NiSe$_2$ is dynamically unstable. Without considering SOC, the A$_1$-band and doubly degenerate E-band of NiSe$_2$ cross about 0.65 eV below the $E_F$, resulting in the symmetry-protected BDP point (red circle in Figure 1e). In particular, the SOC slightly reduces the energy splitting of the E-band, as marked by the red arrow in Figure 1f. Accordingly, the type-II BPD of NiSe$_2$ is located about 0.44 eV below the $E_F$ (red circle in Figure 1f). This result strongly suggests that the SOC in NiTe$_{2-x}$Se$_x$ alloys can be effectively tuned by Se content ($x$), although 1$T$-NiSe$_2$ is unstable. Note that NiSe$_2$ has a pyrite-like structure and crystallizes in the cubic $T_h^6$ space group.[26,27]

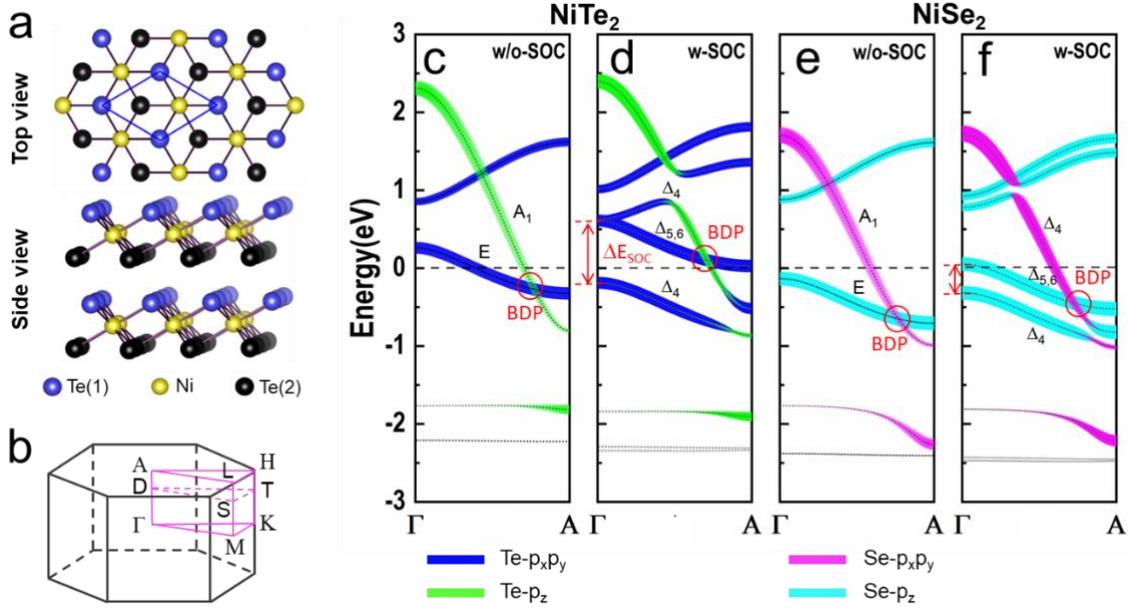

**Figure 1. Orbital-resolved band structures of NiTe₂ and NiSe₂. (a) Top and side views of the 1T-NiTe₂ bulk crystal structure. The unit cell is indicated by the solid blue line. Blue, black, and yellow spheres represent top-Te, bottom-Te, and Ni atoms, respectively. (b) The Brillouin zone of 1T-NiTe₂ with high-symmetry points. The BDP is located at the D-point (0, 0, ±0.35) along the high-symmetry Γ-A direction. (c-f) Orbital-resolved band structures of NiTe₂ and NiSe₂ without (w/o) and with (w) spin-orbit coupling (SOC) along the Γ-A direction.**

In order to demonstrate the tunability of the Dirac band, as predicted by the DFT calculations, high-quality single crystals of NiTe$_{2-x}$Se$_x$ are synthesized with various Se content ($x$ = 0, 0.25, and 1). Figure 2a presents the X-ray diffraction results of NiTe₂, NiTe$_{1.75}$Se$_{0.25}$, and NiTeSe crystals. All peaks are well oriented along the $c$-axis, indicating the single-crystallinity of our samples. The diffraction peaks slightly shift towards a higher angle (i.e., the lattice constant of NiTe$_{2-x}$Se$_x$ decreases) as the Se content ($x$) increases. This is caused by the fact that the atomic radius of Se is smaller than that of Te. The measured lattice constant along the *c*-axis decreases linearly as a function of Se content (as shown in Figure 2b), which agrees well with the DFT calculation results following Vegard's law.

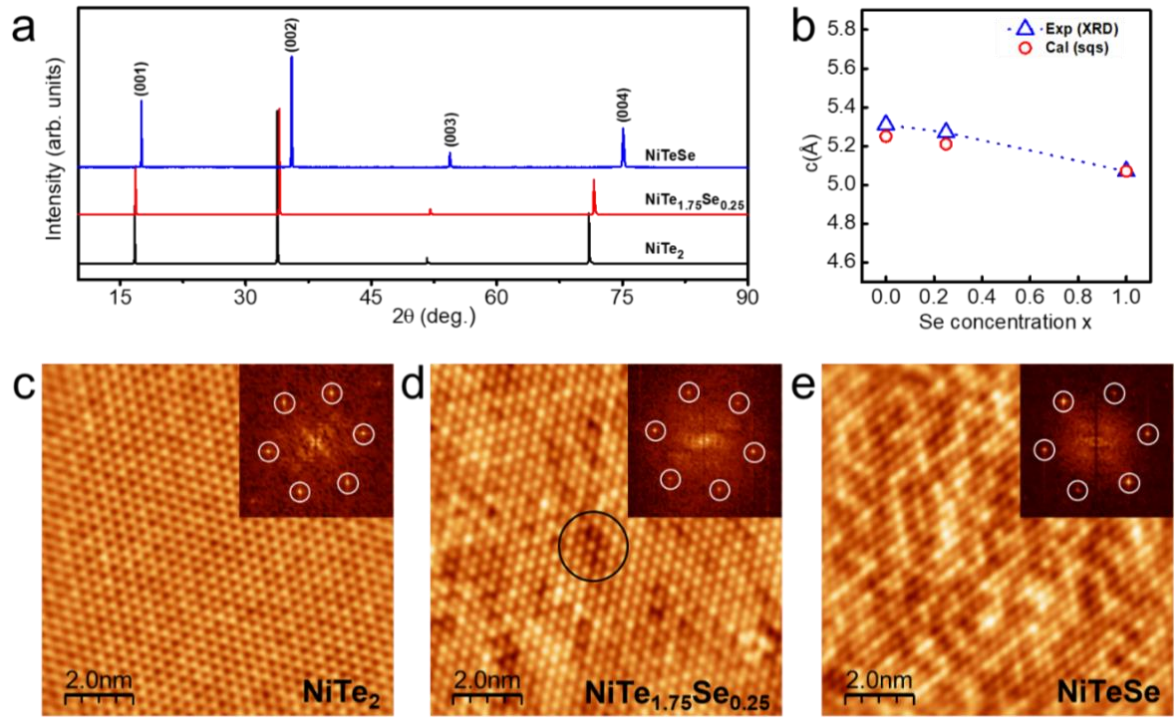

**Figure 2. Characterization of NiTe$_{2-x}$Se$_x$ single crystals. (a) XRD data of NiTe$_{2-x}$Se$_x$ single crystals (*x* = 0, 0.25, 1) on the cleavage plane. (b) The variation of the lattice constant in NiTe$_{2-x}$Se$_x$ along the *c*-axis follows Vegard's law. STM topography and corresponding FFT (inset) images of (c) NiTe$_2$, (d) NiTe$_{1.75}$Se$_{0.25}$, and (e) NiTeSe (*I*$_t$ = 30 pA, *V*$_b$ = 1.0 V). White circles in FFT images indicate the hexagonal lattice.**

The spatial distribution of the Se atoms is explored on the atomic scale with STM. Compared to NiTe$_2$, which presents a clear hexagonal lattice on the top Te layer (Figure 2c), the NiTe$_{1.75}$Se$_{0.25}$ alloy shows a dark feature (marked by a circle), originating from the different atomic sizes of the Se and Te atoms. Accordingly, NiTeSe exhibits more distinct dark (or suppressed intensity) features, which are attributed to highly substituted Se atoms with scattered distribution (Figure 2e). Nonetheless, hexagonal crystallinity is well preserved without significant disorientations in $NiTe_{1.75}Se_{0.25}$ and NiTeSe as shown in the FFT (fast Fourier transformation) images of Figure 2c-e. More topography images of $NiTe_{2-x}Se_x$ at different bias voltages are provided in Figure S3.

To identify the Dirac bands, we measured the photon energy dependent ARPES map and calculated band structures of $NiTe_2$, $NiTe_{1.75}Se_{0.25}$, and NiTeSe. Figure 3a-c show the in-plane ARPES maps of $NiTe_2$, $NiTe_{1.75}Se_{0.25}$, and NiTeSe along the T-D-T direction, respectively (see ARPES maps with a gray scale and corresponding 2D curvature maps in SI Figure S4).[28] The calculated band structures (red shaded lines) are overlaid on the ARPES maps in Figure 3d-f. The ARPES map of $NiTe_2$ and and $NiTe_{1.75}Se_{0.25}$ reproduces the dispersive bands except for some energy offset, possibly due to insufficient handling of the electron correlations in the calculations or probable strain effects at the sample surfaces.[23] Weak intensities of the Dirac bands are present at near $k_x$=0 1/Å, $E_{bin}$=0 eV for $NiTe_2$ and $NiTe_{1.75}Se_{0.25}$ (see 2D curvature maps in Figure S4d and e), respectively.[29] For NiTeSe, on the other hand, a linear-band-crossing point is visible near –0.3 eV below the $E_F$, as marked by the arrow (BDP) in Figure 3c, f. The measured ARPES results of NiTeSe agree with band calculations, except for some energy offset. It is clear that the energy location of BDP moves downward from +0.1 eV in $NiTe_2$ to –0.3 eV in NiTeSe. On the other hand, the photoemission measurement is unable to sufficiently resolve the bands in the energy range -0.6 eV to -1.0 eV. The ARPES intensity is a product of the spectral function and the photoemission matrix elements. The low ARPES intensities of some bands may arise from the small value of the corresponding matrix element for the given measurement geometry and photon energies. The corresponding bands are away from the BDP crossing band that is the focus of the current discussion. Note that bulk DFT calculations do not have the surface state (SS) bands (marked with arrows in Figure 3a-c) associated with the truncated surface effects as reported in previous studies.[22,30] The surface states have a topological origin with chiral spin texture in NiTe$_2$ bulk, while similar surface states have been reported in PtTe$_2$ and PdTe$_2$.[9,31] Since the surface states are independent of photon energies, the in-plane ARPES maps are compared for several photon energies in the Supporting Information (Figure S5).

In addition, Figure 3g-i show the out-of-plane ARPES maps along the Γ-D-A direction overlaid with the band structures (red shaded lines, see also 2D curvature maps in Figure S4). The $k_z$ dispersion of the NiTe$_2$ and NiTe$_{1.75}$Se$_{0.25}$ crystal shows only the bottom part of the Dirac band because the BDP is located above the $E_F$. On the other hand, the $k_z$ dispersion of NiTeSe is renormalized by SOC engineering, and the BDP is located near –0.3 eV below E$_F$, as shown in Figure 3i. Therefore, a similar downshift of BDP is confirmed along the $k_z$ direction as well. It is noted that the out-of-plane ARPES maps show broad features due to the limited $k_z$ resolution of photoelectron, which is similar to the previous reports.[22,24]

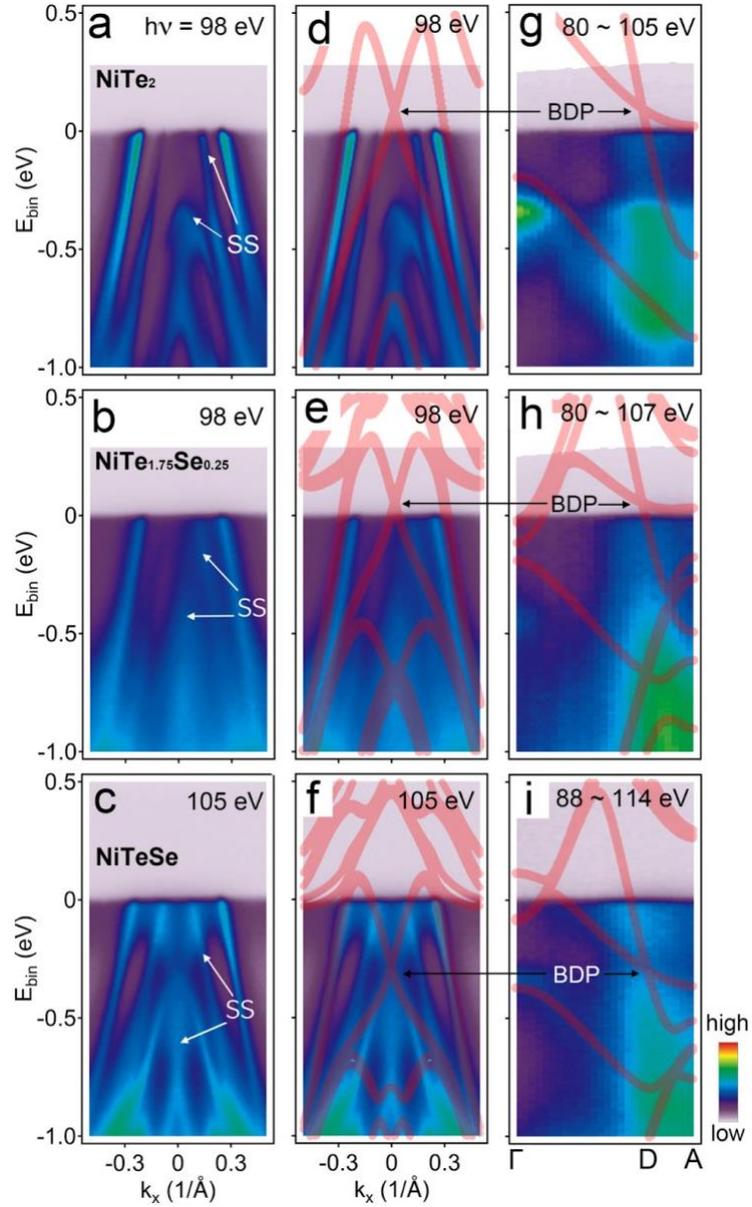

**Figure 3. Comparison of ARPES and DFT results for NiTe₂ (top), NiTe₁.₇₅Se₀.₂₅ (middle), and NiTeSe (bottom). (a-c) in-plane ARPES map along T-D-T. (d-i) ARPES maps and our DFT calculations (red shaded lines) along T-D-T and Γ-D-A. The arrows indicate the BDP and the surface state (SS) bands.**

The comparisons between PDOS calculations and d$I$/d$V$ spectra of NiTe$_{2-x}$Se$_x$ alloy are shown in Figure S6. Although the peak-to-peak comparisons are somewhat limited, our results still show some qualitative agreement. The negative shift of peaks near the Fermi level $E_F$ in STS spectra, reflecting the evolutions of tilted Dirac bands, agrees well with DFT and ARPES results. Discussion details can be found in the Supporting Information.

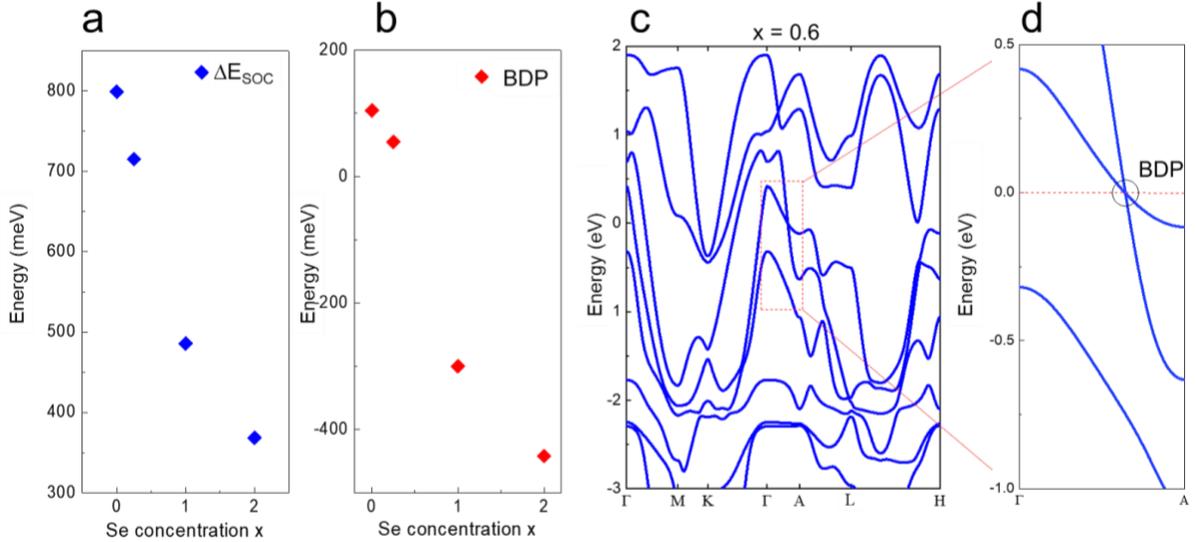

**Figure 4. Controlling the SOC strength and bulk Dirac point (BDP). (a) Modulation of the SOC strength ($\Delta E_{SOC}$) and (b) BDP as a function of the Se concentration in the 1$T$-NiTe$_{2-x}$Se$_x$ alloy.** The SOC strength and BDP are gradually reduced with increasing the Se concentration $x$. **(c) Band structure of 1$T$-NiTe$_{1.4}$Se$_{0.6}$ including SOC. (d) The enlarged band structure shows that the type-II BDP is located at the exact $E_F$.**

From our DFT calculations, we determine the SOC strength and BDP of the NiTe$_{2-x}$Se$_x$ alloy, and find that they gradually decrease with increasing Se content as shown in Figure 4a and 4b. Particularly, the BDP is shifted from +100 meV to –440 meV across the $E_F$ as Te atoms are substitutionally replaced with Se atoms. This result implies that our method to control the BDP via

Se substitutional alloying is as effective as conventional electron doping. Furthermore, our calculations indicate that the type-II BDP is located at the exact $E_F$ for $1T$-NiTe$_{1.4}$Se$_{0.6}$ (see Figure 4c, d), supporting the idea that our proposed approach can facilitate the application of topological transport, such as topological superconductivity and Majorana fermions, anisotropic magnetoresistance, chiral anomaly, 3-dimensional Dirac plasmon.[9,10,21,32–34] On the other hand, the control of precise stoichiometry in ternary compound NiTe$_{2-x}$Se$_x$ is quite tricky due to incongruent melting issues. For instance, selenium easily evaporates during the growth due to its very high vapor pressure, affecting the stoichiometry of the resulting NiTe$_{2-x}$Se$_x$ crystal. Further experimental investigations are required to support our theoretical prediction in Figure 4.

**CONCLUSION**

Tunability of the Dirac band, particularly adjusting the level of Dirac point, has been an important issue in type-II Dirac materials to utilize their relativistic carriers. Here, we focus on the role of chalcogen elements related to SOC control for the purpose of tailoring the type-II Dirac band in NiTe$_{2-x}$Se$_x$ alloys. Theoretical calculations indicate that the SOC strength and BDP are almost linearly tunable, depending on the Se content in the NiTe$_{2-x}$Se$_x$ alloys. The results of the PDOS calculation and STS are in agreement, reflecting the shift of BDP in NiTe$_{2-x}$Se$_x$. ARPES band maps confirm that the BDP significantly moves to –0.3 eV below the $E_F$ in NiTeSe (from +0.1 eV above the $E_F$ in NiTe$_2$). ARPES results further show that the NiTeSe preserves the type-II Dirac fermions with anisotropic Dirac dispersion. As a result, we expect our proposed approach could be the key to effectively controlling the topological character of other chalcogen compounds.

**METHODS**

**Theoretical Calculations**: A $2 \times 2 \times 1$ supercell of bulk $NiTe_2$ was used to study the effects of alloying for Se concentrations of 12.5% and 50%. To model the distribution of the model atoms, we used the special quasi-random structure (SQS) method.[35,36] All the first-principles calculations were carried out based on DFT and implemented in the Vienna Ab-initio Simulation Package (VASP).[37,38] The generalized gradient approximation (GGA)[39] with the optB86b-vdW exchange functional[40,41] (for van der Waals correction) was used for the structural relaxation of $NiTe_{2-x}Se_x$ ($x = 0, 0.25, 1$) alloys. The Perdew-Burke-Ernzerhof (PBE)[39] exchange functional was adopted for all calculations. The projector augmented wave (PAW)[42] method was used to describe ion-electron correlation. Plane waves are included up to the kinetic energy cutoff of 520 eV. The calculations are converged in energy to $10^{-6}$ eV/cell, and the structures are relaxed until the forces are less than $10^{-3}$ eV/Å. Monkhost-Pack k-meshes were used to sample the Brillouin zone.[43] The $\Gamma$-centered $10 \times 10 \times 6$ and $5 \times 5 \times 6$ k-point meshes were used for the pristine $NiTe_2$, $NiSe_2$, and $NiTe_{2-x}Se_x$ alloys, respectively. The electronic properties of $NiTe_{1.4}Se_{0.6}$ were calculated using virtual crystal approximation by mixing the Te and Se PAW potentials.

**Sample preparation**: Single crystals were grown by using the vertical temperature gradient Bridgman method. For the preparation of $NiTe_{2-x}Se_x$ single crystals, we used high-purity Ni (99.999 %), Te (99.99999 %), and Se (99.999 %) elements as starting materials. The elements were put into a quartz ampoule with a capillary bottom in order to grow only one seed after the inside wall of the ampoule was coated with carbon. Then, the ampoule was evacuated and sealed under a pressure of $10^{-6}$ Torr and placed in a furnace. First, the furnace was raised up to 600 ℃ and held for 48 h. The reason for maintaining this intermediate temperature was to avoid excessive pressure buildup due to the high vapor pressures of Te and Se and possible explosion during $NiTe_{2-}$ $_x$Se$_x$ formation. Then, the furnace temperature was raised to 1150 °C at a rate of 20 °C/h, which was followed by soaking for 72 h. For single crystal growth, the temperature was slowly cooled to room temperature over 10 days. The single crystals had a cylindrical form of about 10 mm in diameter and 15 mm in length. The mole fraction $x$ was determined by an electron probe microanalyzer (EPMA-1400, SHIMADZU).

**Scanning Tunneling Microscopy and Spectroscopy Measurements**: All STM and STS measurements were performed using a low-temperature, home-built STM under a base pressure of $\sim 7 \times 10^{-11}$ Torr at 79 K.[44] NiTe$_{2-x}$Se$_x$ single crystals were cleaved *in situ* to obtain clean surfaces for STM measurements. STM topography was acquired in the constant-current mode with the bias voltage applied to the sample. Tungsten tips were electrochemically etched and cleaned *in situ* by electron beam heating. Differential conductance (d$I$/d$V$) spectra were measured by using a standard lock-in technique with a modulation voltage of 10 mV at a frequency of 1.2 kHz. The d$I$/d$V$ curve in Figure S6 is averaged from individual spectra obtained at the $4 \times 4$ grid positions equally distributed over the area of topography images in Figure 2; i.e. total 16 spectra taken at the $4 \times 4$ grid positions are averaged in the d$I$/d$V$ curve.

**Angle-resolved Photoemission Spectroscopy Measurements**: All ARPES measurements were performed in a micro-ARPES end-station (base pressure of $\sim 3 \times 10^{-11}$ Torr) at the MAESTRO facility at beamline 7.0.2 at the Advanced Light Source, Lawrence Berkeley National Laboratory. The ARPES system was equipped with a Scienta R4000 electron analyzer. The lateral size of the synchrotron beam was estimated to be between 30 and 50 μm. NiTe$_{2-x}$Se$_x$ single crystals were cleaved *in situ* at 20 K to obtain clean surfaces. The sample temperature was kept below 20 K during all ARPES measurements. The total energy and angular resolution of our experiments were better than 20 meV and 0.1°, respectively. A series of measurements was made with various photon energies in the range of $80 - 150$ eV.

## ASSOCIATED CONTENT

### Supporting Information

The Supporting Information is available free of charge at…

Orbital resolved and phonon band structure calculations of $NiTe_2$, $NiSe_2$; STM topography images at different bias voltages of $NiTe_{2-x}Se_x$ samples; and comparison of DFT, STS, and ARPES results of $NiTe_{1.75}Se_{0.25}$.

## AUTHOR INFORMATION

### Corresponding Authors

**Younghun Hwang** - Electricity and Electronics and Semiconductor Applications, Ulsan College, Ulsan 44610, Republic of Korea; Email: younghh@uc.ac.kr

**Young Jun Chang** - Department of Physics and Smart Cities, University of Seoul, Seoul 02504, Republic of Korea; orcid.org/0000-0001-5538-0643; Email: yjchang@uos.ac.kr

**Jaekwang Lee** - Department of Physics, Pusan National University, Busan 46241, Republic of Korea; orcid.org/ 0000-0002-7854-4329; Email: jaekwangl@pusan.ac.kr

**Jungdae Kim** - Department of Physics, and EHSRC, University of Ulsan, Ulsan 44610, Republic of Korea; orcid.org/0000-0002-8567-1529; Email: kimjd@ulsan.ac.kr

### Authors

**Nguyen Huu Lam** - Department of Physics, and EHSRC, University of Ulsan, Ulsan 44610, Republic of Korea

**Phuong Lien Nguyen** - Department of Physics, Pusan National University, Busan 46241, Republic of Korea

**Byoung Ki Choi** - Advanced Light Source (ALS), E. O. Lawrence Berkeley National Laboratory, Berkeley, California 94720, United States; Department of Physics and Smart Cities, University of Seoul, Seoul 02504, Republic of Korea; orcid.org/0000-0003-3080-2410

**Trinh Thi Ly** - Department of Physics, and EHSRC, University of Ulsan, Ulsan 44610, Republic of Korea

**Ganbat Duvjir** - Department of Physics, and EHSRC, University of Ulsan, Ulsan 44610, Republic of Korea; orcid.org/0000-0003-3680-9525

**Tae Gyu Rhee** - Department of Physics and Smart Cities, University of Seoul, Seoul 02504, Republic of Korea

**Yong Jin Jo** - Department of Physics, and EHSRC, University of Ulsan, Ulsan 44610, Republic of Korea

**Tae Heon Kim** - Department of Physics, and EHSRC, University of Ulsan, Ulsan 44610, Republic of Korea; orcid.org/0000-0003-4835-0707

**Chris Jozwiak** - Advanced Light Source (ALS), E. O. Lawrence Berkeley National Laboratory, Berkeley, California 94720, United States

**Aaron Bostwick** - Advanced Light Source (ALS), E. O. Lawrence Berkeley National Laboratory, Berkeley, California 94720, United States

**Eli Rotenberg** - Advanced Light Source (ALS), E. O. Lawrence Berkeley National Laboratory, Berkeley, California 94720, United States

**Author Contributions**

**Notes**

The authors declare no competing financial interest.

**ACKNOWLEDGEMENT**

This work was supported by National Research Foundation (NRF) grants funded by the Korean government (Nos. NRF-2018R1D1A1B07050144, NRF-2018R1A2B6004394, NRF-2019R1A6A1A11053838, NRF-2019K1A3A7A09033389, NRF-2019R1I1A3A01063856, NRF-2020R1A2C200373211, NRF-2021R1A6A3A14040322). MOLIT as [Innovative Talent Education Program for Smart City]. The Advanced Light Source is supported by the Director, Office of Science, Office of Basic Energy Sciences, of the U.S. Department of Energy under Contract No. DE-AC02-05CH11231.

**Table of Contents**

Spin-orbit coupling in NiTe$_2$ is controlled by Se substitution, which linearly tunes the energy level of type-II Dirac point in NiTe$_{2-x}$Se$_x$. Energy shift and anisotropic band dispersion of bulk Dirac point are validated by both theoretical calculations and STM/ARPES measurements.

Nguyen Huu Lam, Phuong Lien Nguyen, Byoung Ki Choi, Trinh Thi Ly, Ganbat Duvjir, Tae Gyu Rhee, Yong Jin Jo, Tae Heon Kim, Chris Jozwiak, Aaron Bostwick, Eli Rotenberg, Younghun Hwang[*], Young Jun Chang[*], Jaekwang Lee[*], Jungdae Kim[*]

**Controlling spin-orbit coupling to tailor type-II Dirac bands**

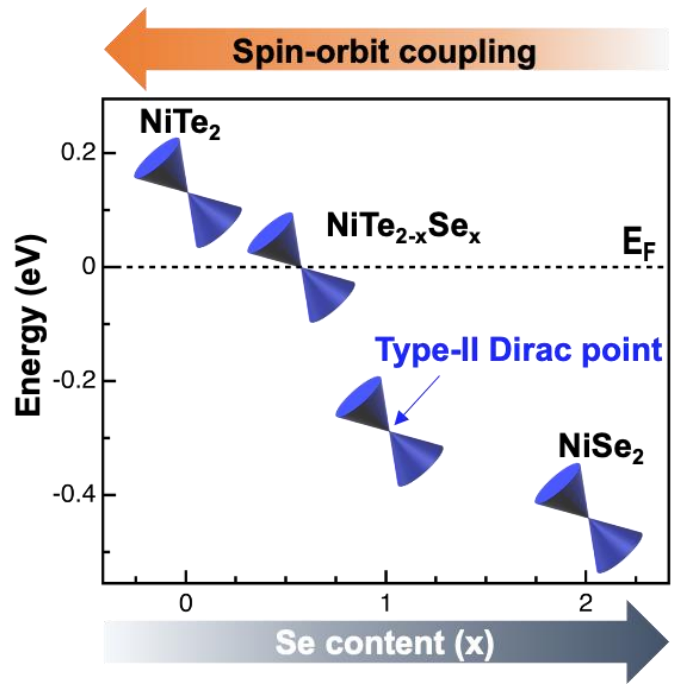